\documentclass[twocolumn,10pt]{article}
\usepackage{latex8}
\usepackage[english]{babel}
\usepackage{epsfig,tabularx,supertabular}
\usepackage{latexsym}
\usepackage{amsmath}
\usepackage{amssymb}
\def\nmr{\hbox{$N$\kern-1pt MR}}

\def\coldot{,} 
 
\def\rightdots#1{%
  \setbox0=\hbox{$1$}\dimen0=#1\wd0
  \setbox0=\hbox{$\coldot$}\advance\dimen0\wd0
  \setbox2=\hbox to \dimen0 {}%
  \setbox0=\hbox\bgroup\mathcode`\.="8000 $}
\def\endrightdots{$\hfil\egroup\box0\box2}
\def\Ariel{{\sc ariel}}
\newcolumntype{d}[1]{>{\hfill\rightdots{#1}}r<{\endrightdots}}
\newcolumntype{D}[1]{>{\hfill\rightdots{#1}}X<{\endrightdots}}
\newcolumntype{C}{>{\centering\arraybackslash}X}
\newcolumntype{L}{X}
\newcolumntype{R}{>{\hfill}X}
\newcommand{\SA}{\hbox{\sc sa}}
\newcommand{\SC}{\hbox{\sc sc}}
\newcommand{\Aty}{\hbox{\sc a}}
\newlength{\widthplusfourty}
\newcommand{\ReL}{\hbox{$\mathcal R\!\raise2pt\hbox{$\varepsilon$}\!\hbox{$\mathcal L$}$}}

\newcommand{\CPP}{$\hbox{C}\hskip-1pt\hbox{+}\hskip-1pt\hbox{+}$}

\begin{document}
\title{$\mathcal R\!\raise2pt\hbox{$\varepsilon$}\!\hbox{$\mathcal L$}$: A Fault Tolerance
       Linguistic Structure\\for Distributed Applications}
\date{}
%
\author{Vincenzo De Florio and Geert Deconinck \\ \ \\
  Katholieke Universiteit Leuven\\
  Electrical Engineering Department, ELECTA Division,\\
  Kasteelpark Arenberg 10, B-3001 Leuven-Heverlee, Belgium\\
  E-mail: \texttt{deflorio@esat.kuleuven.ac.be}}

\maketitle

%
\begin{abstract} 
The embedding of fault tolerance provisions into
the application layer of a programming language is a non-trivial
task that has not found a satisfactory solution yet.
Such a solution is very important, and the lack of
a simple, coherent and effective structuring technique
for fault tolerance has been termed by researchers
in this field as the
``software bottleneck of system development''.
The aim of this paper is 
to report on the current status of a novel
fault tolerance linguistic structure for
distributed applications characterized by
soft real-time requirements. A compliant prototype
architecture is also described.
The key aspect of this structure is that it
allows to decompose the target fault-tolerant application
into three distinct components, respectively responsible
for (1) the functional service, (2) the management
of the fault tolerance provisions, and (3) the
adaptation to the current environmental conditions.
The paper also briefly mentions a few case studies
and preliminary results obtained exercising the
prototype.
\end{abstract}
\Section{Introduction}
\SubSection{Trusting computer services}
Human society more and more expects and relies on the good quality of complex services
supplied by computers: Computer services are becoming more and more \emph{vital},
in the sense that a lack of timely delivery ever more often can have a severe
impact on capitals, the environment, and even human lives. This state of facts is
the consequence
of the tremendous growth in both the complexity and the crucial character of roles
nowadays assigned to computers.
The extent of this process could be hardly foreseen in the ere-days of modern computing:
Those days the main role of computers was basically that of fast solvers
of numerical problems, which made it to some extent acceptable that outages and wrong
results could occur rather often\footnote{This excerpt from
        a report on the ENIAC activity~\cite{Weik61}
        gives an idea of how dependable computers were in 1947:
        ``power line fluctuations and power failures made
        continuous operation directly off transformer mains
        an impossibility [\ldots]
        down times were long; error-free running periods were short [\ldots]''.
        After many considerable improvements, still
        ``trouble-free operating time remained at about 100 hours a week during
        the last 6 years of the ENIAC's use'', i.e., a reliability
        of about 60\%!}.
Computer failures were a bothering fact to accept and live peacefully with.
The very same increase in computer reliability and performance pushed up the
introduction of computer services till they actually \emph{permeated\/} our society.
Consequently,
what we call the \emph{criticality\/} of computer services---that is, the magnitude
of the consequences of a computer failure---has dramatically increased and, with it,
the need for guarantees that computer failures can be avoided or their extent bounded.
\emph{Dependability}, or the trustworthiness of a computer system
such that reliance can justifiably be placed on the service
it delivers~\cite{Lapr85}, became a fundamental requirement.

Devising methods to fulfil the requirement for dependability of computer services
has been and still is a hot research topic. We are not going to review those methods,
but merely observe that they can be classified according to the (physical or virtual)
machine they address: as an example, hardware fault tolerance (HFT) is the name of
the class of methods that target physical faults and aim at preventing that they
bring the physical machine to a failure. We believe HFT is an important requirement
to achieve a truly dependable computer service, as it addresses the basement
of the hierarchy of machines that collectively supply that service.
Likewise we are convinced that,
as any computer service is the result of the concurrent progress of a hierarchy of 
machines, service dependability may be best reached through a strategy that
target the whole of the hierarchy: Failing to consider a tassel means
weaking a link in the chain---a single point of overall service failure.

The top of the hierarchy---the application layer---is no exception. On the contrary,
a design fault at this level may well be as jeopardizing as a physical fault
in the hardware machine, for the application layer is the very ``place''
where the service is specified (in its more abstract terms).

It is this general purpose character that makes so difficult devising an
application level fault tolerance (ALFT) strategy:
Indeed, while effective solutions have been found, e.g., for the hardware, the
operating system, and the middleware layers, the problem
of \emph{an effective system structure for expressing fault tolerance
provisions in the application layer of computer programs\/} is
still an open one.

Structuring techniques provide means to control complexity,
the latter being a relevant factor for preventing the introduction
of design faults. This fact and the ever increasing
complexity of today's distributed software justify
the need for simple, coherent, and effective
structures for the expression of fault tolerance in the
application software. This paper describes the
``recovery language approach'' (\ReL), i.e.,
a structuring technique for the expression
of the fault tolerance design aspects in the
applications characterized by soft real-time requirements.
The \ReL{} technique in particular addresses three requirements of
fault-tolerant software design:
\begin{description}
\item[R1] Separation of the functional and fault tolerance design aspects,
      such that the two design concerns do not conflict with each other.
\item[R2] Dynamic adaptability to varying environmental conditions,
      obtained through a sort of dynamic linking of the
      fault tolerance executable code.
\item[R3] A syntactical structure capable of hosting
      a wide class of fault tolerance (FT) provisions\footnote{By ``FT provision''
      we mean any strategy (e.g. recovery blocks),
      or mechanism (such as watchdog timers), that can
      be used to introduce FT aspects into an application.}.
\end{description}
The above requirements are met by exploiting \ReL{}'s capability
to partition the design complexity of a distributed application
into three components:
\begin{enumerate}
\item An application-specific component realizing the functional
      specification.
\item A special-purpose component dealing with the management
      of the FT provisions.
\item A special-purpose component responsible for the
      run-time adaptation of the FT provisions
      to the current environmental conditions.
\end{enumerate}

The structure of this paper is as follows:
Section~\ref{s:ReL} introduces the elements of our
approach. Section~\ref{s:ariel} describes a
\ReL-compliant prototype software architecture that
has been developed in the framework of the two ESPRIT
projects EFTOS (``embedded fault-tolerant
supercomputing'')~\cite{DVBD97} and TIRAN (``tailorable
fault tolerance frameworks for embedded applications'')~\cite{BDDC99+}.
That architecture focuses on component 2.
Section~\ref{s:ariel} also mentions
a few case studies where \ReL{} is proving
its effectiveness. The paper is concluded by
Sect.~\ref{s:end}, which also provides the reader
with the elements of a new \ReL-compliant architecture.
Such architecture, which is being developed in the framework
of the IST-2000-25434 project
DepAuDE (``\emph{Dep\/}endability for embedded
\emph{Au\/}tomation systems in
\emph{D\/}ynamic \emph{E\/}nvironments with
intra-site and inter-site distribution aspects''),
is to fully exploit the capabilities of \ReL. The key goal of
this architecture is to realize all the special-purpose
components of a fully \ReL{}-compliant distributed architecture,
leaving to the
user the sole management of the 
service specification.
\Section{The Recovery Language Approach}\label{s:ReL}
This section describes \ReL, a FT
linguistic structuring technique for distributed applications
with soft real-time constraints. By structuring technique
we mean a set of methods by means of which it is possible
to \textbf{express}
and to \textbf{manage}
some FT provision. In the following, we will
characterize both the above ``methods''---expressing
and managing a FT provision. Furthermore, in order
to characterize our technique with respect to the
existing ones, we will make use, informally, of a 
``base'' of structural properties, namely
\begin{description}
\item[\SC:] separation of design concerns,
\item[\Aty:] adaptability to a varying environment, and
\item[\SA:] syntactical adequacy, i.e., the adequacy of the
 technique at hosting a FT provision, averaged on
 the set of possible FT provisions.
\end{description}
Clearly the above properties respectively match
requirement \textbf{R1}, \textbf{R2} and \textbf{R3}.
In what follows we will show that \ReL{} is a simple,
coherent, and effective FT linguistic structure
that provides satisfactory values of the three
structural properties (\SC, \Aty, \SA) in the
domain of soft real-time, distributed applications.

In \ReL{} two distinct programming languages are available
to the programmer:
a service language, i.e., the
programming language addressing the functional design concerns,
and a special-purpose linguistic structure
(called ``recovery language'') for the expression of
error recovery and reconfiguration tasks.
This recovery language comes into play either asynchronously,
as soon as an error is detected by an underlying error detection
layer, or when some erroneous condition is signaled
by the application processes. Error recovery and
reconfiguration are specified as a set of
\emph{guarded actions}, i.e., actions that require a
pre-condition to be fulfilled in order to be executed.
Recovery actions deal with coarse-grained entities
of the application and the system, and pre-conditions query the
current state of those entities. An example of
a recovery action is the following one:
\begin{center}
  \begin{tabbing}
  \textbf{if}  \=  \kill
  \textbf{if}  \> a transient faults affects ``task 10'' :\\
                 \> \textbf{restart} task 10 \\
                 \> \textbf{notify}  the group of tasks to which task 10 belongs\\
  \textbf{end}
  \end{tabbing}
\end{center}
A larger example of guards and actions can be seen in
Sect.~\ref{s:ariel}, where a prototype \ReL-compliant
architecture is described.

An important added value of \ReL{} is that it allows for
the \textbf{expression} of the recovery actions to be
done in a design and programming context other than the
one in which
the expression of the functional service takes place.
This minimizes non-functional code intrusion and hence
enhances property \SC.

\begin{figure}[t]
\centerline{\psfig{figure=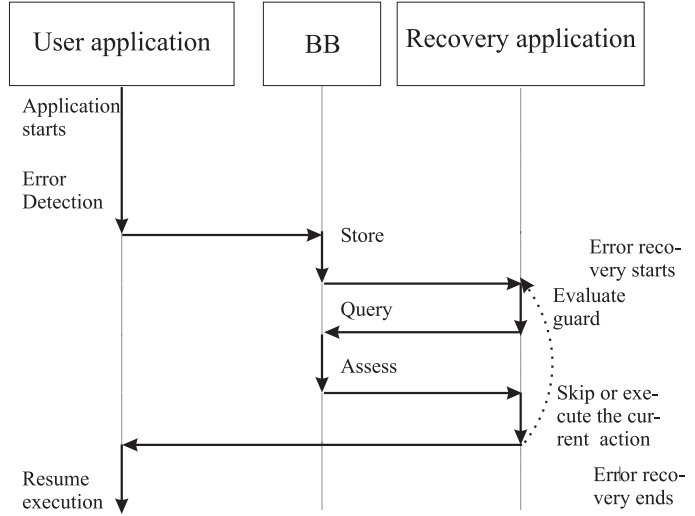,width=9.0cm}}
\caption{Scheme of execution of a \ReL-compliant
 application: together with the application, two special-purpose
 tasks are running---a system-wide database management system
 (we call it the ``backbone''),
 which stores error detection notifications sent by a periphery
 of detection tools, and a ``recovery application'', i.e.,
 a task responsible for the execution of the recovery actions.
 The diagram describes the execution of the
 user-specified recovery actions. The dotted line represents
 a jump to the execution of the next guarded action, if any.
 Error recovery ends when the last guarded action is evaluated.}
\label{f:ua-db-ra}
\end{figure}

The execution of the recovery actions is done via
a fixed (i.e., special-purpose) scheme, portrayed in
the sequence diagram of Fig.~\ref{f:ua-db-ra}: as soon as
an error is detected, a notification describing that event
is sent to a distributed entity responsible 
for the collection and the management of these notifications.
Let us call such entity the ``backbone'' (BB). Immediately
after storing each notification, the guards of the
recovery actions are evaluated. Guards evaluation
is done by querying the BB. When a guard is
found to be true, its corresponding actions are executed, 
otherwise they are skipped.

The just sketched strategy represents the way \ReL{} 
performs its \textbf{management} of the FT provisions 
to be embedded in the target application. An important
consequence of the adoption of this strategy is that
the functional executable code and the non-functional 
executable code are \emph{distinct}: the former implements
the user tasks, while the latter is given by a proper
coding of the recovery actions.
This allows to decompose the design process into two
distinct phases. When the interface between the two
``aspects'' is simple and well-defined, this provides
a way to control the design complexity, which decreases development
times and costs.
In the current implementation,
described in Sect.~\ref{s:ariel}, the recovery
actions are translated into a ``recovery pseudo-code''
(we call it r-code) that is interpreted by an
r-code virtual machine. Currently, the r-code can 
either be read from a file or ``hardwired'' in the r-code
virtual machine. The separability of the r-code
from the functional code provides the elements
for the approach described in Sect.~\ref{s:end},
which focuses on adaptability and FT software reuse.

The above strategy clearly focuses on the
\emph{error recovery\/} step of FT. In order to minimize the code intrusion
due to \emph{error detection\/} and \emph{fault masking},
we envisaged a \textbf{configuration language} that allows 
the user to set up ready-to-use instances of provisions 
selected from a custom library of single-version FT mechanisms,
including, e.g., a watchdog timer or a voting tool.
These instances are also instrumented in such a way as
to forward transparently their notifications to the BB.
Notifications include, e.g., a watchdog timer's alarm, or
a caught division-by-zero exception, or
a minority input value to a voting tool. An example of configuration
language can be seen in Sect.~\ref{s:ariel}. The same
translator that turns the recovery actions into the r-code
is used in that case to write the source files with the 
configured instances.

\SubSection{System and Application Models}
The target system for \ReL{} is assumed to be
a distributed or parallel system. Basic
components are nodes, tasks, and the network.
A node can be, e.g.,
a workstation in a networked cluster or a processor
in a MIMD parallel computer.
Tasks are independent
threads of execution running on the nodes.
The network system allows tasks on different nodes
to communicate with each other.
Nodes can be commercial-off-the-shelf hardware
components with no special provisions for hardware
FT.
A general-purpose operating system (OS) is required on
each node.
No special purpose, distributed, or fault-tolerant OS is
required.
The system obeys the \emph{timed asynchronous distributed system
model\/}~\cite{CrFe99}:\label{tads}
\begin{itemize}
\item Tasks communicate through the network via a
      datagram service with omission/per\-for\-mance failure semantics~\cite{Cri91}.
\item Services are timed: specifications prescribe not only
      the outputs and state transitions that should occur
      in response to inputs, but also the time intervals
      within which a client task can expect these outputs and
      transitions to occur.
\item Tasks (including those related to the OS and the network)
      have crash/performance failure semantics~\cite{Cri91}.
\item Tasks have access to a node-local hardware clock.
      If more than one node is present, clocks on different nodes
      have a bounded drift rate.
\item A ``time-out'' service is available at application-level:
      using it, tasks can schedule the execution of events so that
      they occur at a given future point in time, as
      measured by their local clock.
\end{itemize}
In particular, this model allows a straightforward modeling
of system partitioning---as a consequence of sufficiently many
omission or performance communication failures, correct nodes
may be temporarily disconnected from the rest of the system
during so-called periods of instability~\cite{CrFe99}.
A message passing library is assumed to be available, built on the
datagram service. Such library offers asynchronous, non-blocking
multicast primitives.
As clearly explained in~\cite{CrFe99}, the above
hypotheses match well to nowadays distributed systems based
on networked workstations---as such, they represent a general
model with no practical restriction.
The following assumptions characterize the
user application:
\begin{itemize}
\item The service is supplied by a distributed application.
\item It is written or is to be written in a procedural 
      or object-oriented language such as C or Java.
\item The application is non safety-critical.
\item The target application is characterized by soft
      real-time requirements. In particular, performance
      failures may occasionally show up during error recovery.
\item Inter-process communication takes place
      by means of the functions in the above mentioned
      message passing library. Higher-level communication
      services, if available, must be based on the message passing
      library as well.
\end{itemize}
As suggested, e.g., in~\cite{SaRC84},
any effective design including dependability goals
requires provisions, \emph{located at all levels},
to avoid, remove, or tolerate faults.
Hence, as an \emph{application-level\/} structuring technique,
\ReL{} is complementary to other approaches addressing
FT at \emph{system level}, i.e.,
hardware-level and OS-level FT.
In particular, a system-level architecture such
as GUARDS~\cite{Pow99}, that is based on redundancy and
hardware and OS provisions for systematic management
of consensus, appears to be particularly appropriate
for being coupled with \ReL{} which offers application-level
provisions for N-version programming and replication (see Sect.~\ref{s:ariel}).

\SubSection{Work-flow of \ReL}\label{ss:wf}
This section describes the work-flow corresponding to the
adoption of the \ReL{} approach.
\begin{figure}[t]
\centerline{\psfig{figure=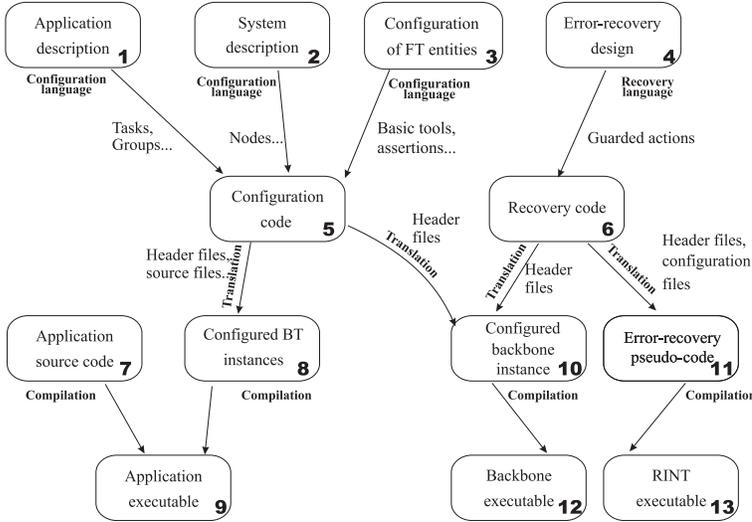,width=10.0cm}}
\caption{A work-flow diagram for \ReL. Labels refer to usage steps
and are described in Sect.~\ref{ss:wf}.}
\label{Fig:WF}
\end{figure}
Figure~\ref{Fig:WF} summarizes the work-flow.
The following basic steps have been foreseen:
\begin{itemize}
\item In the first steps (labels 1 and 2 in the cited figure),
      the designer describes the key
      application and system entities, such as tasks,
      groups of tasks, and nodes. The main tool
      for this phase is the configuration language.
\item Next (step 3), the designer configures a number
      of basic FT tools (BTs)
      he or she has
      decided to use. The configuration language
      is used for this. The output of steps 1--3
      is the configuration code.
\item Next (step 4), the designer defines which conditions need
      to be caught, and which actions should follow
      each caught condition. The resulting list is coded
      as a number of guarded actions via a
      recovery language.
\item The configuration code and the recovery code are
      then converted via the translator into a
      set of C header files, C fragments, and
      system-specific configuration files (steps 5
      and 6).
      These files represent: configured instances
      of the BTs, of the system and of the
      application; initialization files
      for the communication management functions;
      user preferences for the BB;
      and the recovery pseudo-code.
\item On steps 7--9, the application source code
      and a set of configured instances of BTs
      are compiled
      in order to produce the executable codes
      of the application.
\item Next, the BB and the recovery interpreter are compiled
      on steps 10--13.
\end{itemize}
The resulting components, i.e., the executable codes of
the application, the backbone, and RINT,
represent the entities portrayed in Fig.~\ref{Fig:WF}.

In the following we briefly summarize the specific differences
between ours and other novel approaches.

\SubSection{Specific Differences with respect to Other Approaches}
Numerous techniques have been devised in the past to solve the problem
of optimal and flexible development of dependability services to be embedded
in the application layer of a computer program.
In~\cite{DF00}, some of these approaches are critically reviewed and qualitatively assessed
with respect to a set of structural attributes (separation of design concerns,
syntactical adequacy and adaptability).
A non-exhaustive list of the systems and projects implementing 
these approaches is also given in the cited reference.
In particular, approaches based on metaobject protocols~\cite{KirB91} (MOPs),
FT distributed programming languages~\cite{Rob99}
and aspect-oriented programming~\cite{KLM97} (AOP)
are reviewed therein. 

\paragraph{Metaobject Protocols.}\label{ss:mops}
The key idea behind MOPs is that of ``opening'' the implementation of the
run-time executive of an object-oriented language like C++ or Java so that
the developer can adopt and program different, custom semantics,
adjusting the language to the needs of the user and to the requirements
of the environment.
Using MOPs, the programmer can modify the behavior of fundamental features like
methods invocation, object creation and destruction, and
member access. 
The key concept behind MOPs is that of \emph{computational reflection}, or the
causal connection between a system and a meta-level description
representing structural and computational aspects of that system~\cite{Maes87}.
An architecture supporting this approach is FRIENDS~\cite{FaPe98}.
FRIENDS implemented a number of FT provisions (e.g., replication, group-based
communication, synchronization, voting) as MOPs.

A number of studies confirm that MOPs
reach efficiency in some cases~\cite{KirB91}, though no experimental
or analytical evidence allows to estimate the practicality and
the applicability of this approach~\cite{RaXu95,LiLo00}.
MOPs only support object-oriented programming
languages and require special extensions or custom programming languages.
\paragraph{Aspect-oriented Programming Languages.}\label{ss:aop}
Aspect-oriented programming~\cite{KLM97} is a
programming methodology and a structuring technique
that explicitly addresses,
at system-wide level, the problem of the best code structure
to express different, possibly conflicting design goals
like for instance high performance, optimal memory usage, or
dependability.

Developed as a Xerox PARC
project, AspectJ is an aspect-oriented
extension to the Java programming language~\cite{Kic00,LiLo00}.
A study has been carried out on the
capability of AspectJ as an AOP language supporting
exception detection and handling~\cite{LiLo00}. It has been
shown how AspectJ can be used to develop so-called
``plug-and-play'' exception handlers: libraries of
exception handlers that can be plugged into many
different applications. This translates into better
support for managing different configurations
at compile-time. Up to now, no AOP tool or programming
language exists for flexible development of
{\em dependable\/} services:
AspectJ only addresses exception detection
   and handling. Remarkably enough, the authors of
   a recent study on AspectJ and its support to this field
   conclude~\cite{LiLo00} that ``whether the properties of
   AspectJ [documented in this paper] lead to programs
   with fewer implementation errors and that can be changed easier,
   is still an open research topic that will require serious
   usability studies as AOP matures''.
\Section{The \Ariel{} Configuration and Recovery Language}\label{s:ariel}
This section describes a prototypic architecture based on \ReL{}
that has been developed during recently ended project TIRAN.
In the following, in Sect.~\ref{s:tiran} we present the contents of
TIRAN. The main components of the
TIRAN architecture are then briefly introduced in Sect.~\ref{s:t:fw}.
In particular, the TIRAN recovery language, \Ariel{},
is reported in Sect.~\ref{s:t:rl} and a few case studies 
in Sect.~\ref{s:case}.
\SubSection{The TIRAN Project}\label{s:tiran}
The main objective of project TIRAN (ESPRIT 28620) has been to develop
a software framework that provides
fault-tolerant capabilities to automation systems.
Application-level support to FT is provided by means
of a \ReL{}-compliant architecture, which is described 
in the rest of this section.
The framework provides a library of software FT provisions
that are parametric and support an easy configuration process.
Using the framework, application developers are allowed to
select, configure and integrate provisions
for fault masking, error detection, isolation and recovery among those
offered by the library.
Goal of the project is to provide a tool that
significantly reduces the
development times and costs of a new dependable system.
The target market segment concerns
non-safety-critical distributed soft-real-time embedded
systems~\cite{BDDC99b}.
TIRAN explicitly adopts formal techniques to support requirement specification and
predictive evaluation~\cite{DoBo00}. This, together with the intensive testing on pilot
applications, is exploited in order to:
\begin{itemize}
\item Assess the correctness of the framework.
\item Quantify the fulfillment of time,
      dependability and cost requirements.
\item Provide guidelines to the configuration process of the users.
\end{itemize}
Most of this framework has been designed for being
platform independent. A single version of the framework
has been written in the C programming language making use
of a library of ``basic services'' (BSL) developed by the
TIRAN consortium. The TIRAN framework is currently
running on Windows-NT, Windows-CE, the Virtuoso microkernel~\cite{Vir98},
VxWorks, and the TEX microkernel~\cite{Anon97b}. 

The project results, driven by industrial users' requirements and
market demand, is being integrated into the Virtuoso microkernel
and adopted by ENEL and SIEMENS within their application fields.

\begin{figure}[t]
  \centerline{\psfig{figure=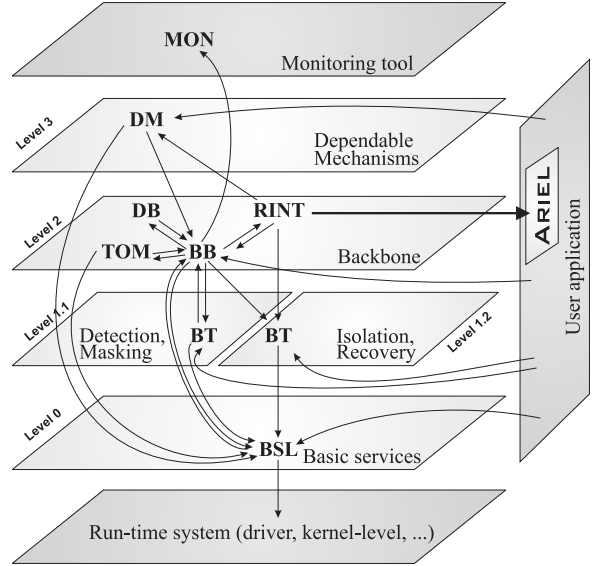,width=8cm}}
  \caption{A representation of the TIRAN elements.
    The central, whiter layers constitute the TIRAN framework.
    This same structure is replicated on each processing node of
    the system.}
  \label{Fig:TIRANstruct}
\end{figure}

\SubSection{The TIRAN Framework}\label{s:t:fw}
Figure~\ref{Fig:TIRANstruct} draws the TIRAN architecture and
positions its main components into it. In particular, the
box labeled ``Ariel'' represents the TIRAN recovery
language, \Ariel.
The central, whiter layers represent the TIRAN framework.
In particular:
\begin{itemize}
\item Level 0 hosts
      the BSL (see Sect.~\ref{s:ariel}), which gives system-independent access to the
      services provided by the underlying run-time system.
\item Level 1 services are provided by a set of BTs
      for error detection and fault masking (level 1.1)
      and by another set addressing isolation, recovery
      and reconfiguration
      (level 1.2). These services are not distributed on multiple nodes.
\item Level 2 hosts the TIRAN BB~\cite{DeDL00d}.
      This is the component responsible for the management of the 
      distributed database (DB) that maintains records describing
      errors detected by Level 1.1 BTs.
      It also includes a time-out management system, called TOM~\cite{DeDL00a}, and a
      recovery interpreter, RINT, actually a
      virtual machine executing the r-code.
      The BB executes an algorithm, described in~\cite{DeDL00d},
      which allows it to tolerate node and component crashes and to
      withstand partitioning caused by temporary periods of communication
      instability. The BB straightforwardly supports the $\alpha$-count 
      fault identification mechanism~\cite{BCDG00} by feeding $\alpha$-count
      filters immediately after the arrival of each new error detection
      notification.
      In Fig.~\ref{Fig:TIRANstruct}, the edge
      connecting RINT to \Ariel{} means that RINT
      actually implements (executes) the \Ariel{} programs.
      Note the control and data messages that flow from
      BB to TOM, DB, and RINT.
      RINT also sends control messages to the isolation
      and recovery BTs. These are low-level messages
      that request specific recovery actions.
      Data messages flow also from BB to a monitoring tool~\cite{DeDe98}.
\item Dependable mechanisms (DMs), i.e., high-level, distributed
      FT tools exploiting the services of the BB and of the
      BTs, are located at level 3. These tools include a distributed
      voting tool~\cite{DeDL98e}, a distributed synchronization tool, 
      and a data stabilizer~\cite{PDP2001}.
      The DMs receive notifications from RINT in order to execute
      reconfigurations such as, for instance, introducing a spare task
      to take over the role of a failed task.
\end{itemize}
The layers around the TIRAN framework in Fig.~\ref{Fig:TIRANstruct}
represent (from the layer at the bottom and proceeding counter-clockwise):
\begin{itemize}
\item The run-time system.
\item The functional application layer and the recovery language
      application layer (again, box labeled ``Ariel'').
\item A monitoring tool, for hypermedia rendering of the current state
      of the system within the windows of a WWW browser.
\end{itemize}

Next section focuses on the key component of the TIRAN
prototype, namely, the \Ariel{} recovery language.
\SubSection{The \Ariel{} Language}\label{s:t:rl} 
Within TIRAN, a single syntactical framework---provided by
the \Ariel{} language---serves the application designer
as both a configuration and a recovery language. 
\Ariel{} is a language with a syntax
somewhat similar to that of the UNIX shells.
\Ariel{} deals with
five basic types:
``\emph{nodes\/}'', ``\emph{tasks\/}'',
``\emph{groups\/}'', integers, and real numbers.
A node is a uniquely identifiable processing node
of the system, e.g., a processor of a MIMD
supercomputer. A task is a uniquely identifiable
process or thread in the system. A group is a
uniquely identifiable collection of tasks,
possibly running on different nodes.
Nodes, tasks, and groups are generically called
\emph{entities}. Entities are uniquely
identified via non-negative integers; for
instance, \texttt{NODE3} or \texttt{N3}
refer to processing node currently configured
as number 3.
Symbolic constants can be ``imported'' from C language
header files through the statement \texttt{INCLUDE}.
When curly brackets appear around a string, the value
of the corresponding symbolic constant is returned.

The key statement in \Ariel{} is the \texttt{IF},
which is used to code a recovery action as follows:
\begin{center}
\texttt{IF [} \emph{guard\/} \texttt{] THEN} \ \emph{actions},
\end{center}
where a \emph{guard} checks whether an entity, according to
the current contents of the database, is in one
of the following states: active; affected by a fault; affected by
a transient fault; isolated; restarted. A guard can also
check the current ``phase'' of a task, e.g., its current
algorithmic step, that the task can declare via a custom
BSL function. Actions can be guards---which allows to
represent recovery actions as trees---and remote or
local commands for: sending messages to tasks and groups;
terminating, isolating, starting or restarting an entity.
Restarting a node means rebooting it, terminating a
node means performing a node shutdown. Isolating a task
means disabling its communication descriptors. A local
command is executed by the local BB component, while a
remote one is first sent to the corresponding BB component
and then executed by it.

\Ariel{} allows also to configure its BTs. For instance, the
following syntax:
\begin{tt}
\begin{tabbing}
xxx \= xxx \= xxx\kill
     \> INCLUDE "mydefinitions.h"\\
     \> WATCHDOG \{MYWD\} WATCHES TASK \{MYTASK\}\\
     \> \>   HEARTBEATS EVERY \{HEARTBEAT\} MS\\
     \> \>   ON ERROR WARN TASK \{CONTROLLER\}\\
     \> END WATCHDOG
\end{tabbing}
\end{tt}
produces a source code configuring a watchdog that, once
enabled by its first heartbeat, expects new such
messages every \texttt{HEARTBEAT} milliseconds,
or sends task \texttt{CONTROLLER} an alarm message.
Note that in this case the error detection
code intrusion is reduced to the function call
for sending heartbeats.
Configuration also includes replicated tasks
and N-version programming. Syntaxes for
retry blocks and consensus recovery blocks have
been also implemented.

The \Ariel{} translator, called ``art'', produces
both the configured instances of the BTs and the
recovery pseudo-code (r-code). The latter can either
be output as a binary file, to be read by RINT at run-time,
or as an include file to be compiled with RINT.
This r-code is then re-executed by RINT each time the
backbone notifies it that a new event has been stored
in the database---as described in Fig.~\ref{f:ua-db-ra}.
\SubSection{Case Studies}\label{s:case}
The \Ariel{} language and the TIRAN framework have been
exercised in the course of project EFTOS and project TIRAN on
a number of case studies, in as different an application
domain as postal automation, electrical substation automation,
and airport lighting systems. These case studies
were formulated by two members of the EFTOS and TIRAN
consortia (Siemens and ENEL) and have their
origin within the internal strategies of those
companies. One of these case studies is reported
in~\cite{DeDe02c}.
Another noteworthy case study has been
the development of a Level 3 FT mechanism supporting
distributed voting. This tool exploits two
features of \Ariel: first, it makes use of spare 
components---error recovery
strategies like reconfiguration and graceful
degradation (when spares are exhausted) can be
expressed in terms of \Ariel{} scripts and
result in no code intrusion. Secondly,
it exploits the built-in support of the
$\alpha$-count fault identification mechanism
in order to let the user express different
error recovery strategies depending on the
nature of the corresponding faults. This
allows to express recovery actions
such as:
\begin{small}
\begin{tt}
\begin{tabbing}
xx \= IF \= THEN \= xxx \= \kill
\>IF [ FAULTY TASK \{MYTASK\} ] \\
\>THEN\\
\>\> IF [ TRANSIENT TASK \{MYTASK\} ]\\
\>\> THEN \emph{Conservative strategy}\\
\>\>   \> \emph{(e.g., restart the task)}  \\
\>\> ELSE \emph{Reconfiguration}  \\
\>\> FI\\
\>FI\hbox{.}
\end{tabbing}
\end{tt}
\end{small}
This aims at keeping reconfiguration as the
ultimate solution in order to minimize
the rate at which redundancy is ``consumed''.
Markov modeling of this approach shows that it 
allows to enhance considerably reliability~\cite{DF00}.
For the sake of brevity we refer to the cited sources for a
full description of the case studies and their evaluation.
\Section{Conclusions and Future Work}\label{s:end}
A novel fault tolerance linguistic structure
for distributed applications has been briefly described.
Such structure is at the core of the strategy
that is currently being designed within IST-2000-25434 Project ``DepAuDE''
to allow dependable real-time applications with intra-site
and inter-site distribution aspects to adapt to a changing
environment (\cite{DeDe02b} briefly mentions the key ideas behind
the DepAuDE strategy).
The design of the elements of the architecture sketched in
this paper, which explicitly addresses requirement \textbf{R1},
\textbf{R2} and \textbf{R3}, is one of the goals of DepAuDE.
As mentioned before, \ReL{} is being used in several case studies
with promising results. One of these case studies is described
in~\cite{DeDe02c}.
The adoption of a recovery language within a generative
communication infrastructure (such as the one of LINDA~\cite{Ge85})
is also currently being experimented~\cite{DeDe01}.

\paragraph{Acknowledgements.}
This project is partly supported by
the IST-2000-25434 Project ``DepAuDE''.
Geert Deconinck is a Postdoctoral Fellow of the
Fund for Scientific Research - Flanders (Belgium) (FWO).

\bibliographystyle{latex8}

\end{document}